# SPATIO-TEMPORAL QUERIES FOR MOVING OBJECTS DATA WAREHOUSING

[1]Leila Esheiba, Hoda M.O.Mokhtar[2], and Mohamed El-Sharkawi[3]

[1]Department of Information Systems, Faculty of Computers and Information,
Cairo University, Egypt
`laila.abdelrahman@fci-cu.edu.eg`

[2]Department of Information Systems, Faculty of Computers and Information,
Cairo University, Egypt
`h.mokhtar@fci-cu.edu.eg`

[3]Department of Information Systems, Faculty of Computers and Information,
Cairo University, Egypt
`m.elsharkawi@fci-cu.edu.eg`

*ABSTRACT*

*In the last decade, Moving Object Databases (MODs) have attracted a lot of attention from researchers. Several research works were conducted to extend traditional database techniques to accommodate the new requirements imposed by the continuous change in location information of moving objects. Managing, querying, storing, and mining moving objects were the key research directions. This extensive interest in moving objects is a natural consequence of the recent ubiquitous location-aware devices, such as PDAs, mobile phones, etc., as well as the variety of information that can be extracted from such new databases. In this paper we propose a Spatio-Temporal data warehousing (STDW) for efficiently querying location information of moving objects. The proposed schema introduces new measures like direction majority and other direction-based measures that enhance the decision making based on location information.*

*KEYWORDS*

*Moving objects Databases (MODs), Spatio-Temporal data warehousing (STDW), Moving objects.*

## 1. INTRODUCTION

The rapid advances of wireless, communication systems and the vast advances in technologies for tracking the positions of continuously moving objects are among the reasons for new emerging applications like traffic control, location-based services, fleet management, m-commerce, E-911, and many others. Moving objects are geometries that change their position and/or shape continuously over time. This unique characteristic of moving objects makes traditional database management systems no longer adequate to accommodate moving objects. This inadequacy is due to the fact that traditional databases mainly assume data to be constant unless it is explicitly modified[1]. In order to manage, maintain, access, and query the large amount of location data, new database techniques are needed. In addition, taking critical decisions based on this huge amount of data is a complex procedure. Data warehousing techniques are among the techniques that were introduced to support the decision making process. According to Inmon in[2], a data

    1

International Journal of Database Management Systems ( IJDMS ) Vol.5, No.3, June 2013International Journal of Database Management Systems ( IJDMS ) Vol.5, No.3, June 2013 warehouse (DW) is defined as a subject oriented, integrated, time-variant, non-volatile collection of data that helps and supports the decision making process. Typically, the data warehouse is maintained separately from the organization's operational databases. There are many reasons for doing this, the key reason is that the data warehouse supports on-line analytical processing (OLAP), the functional and performance requirements of which are quite different from those of the on-line transaction processing (OLTP) applications traditionally supported by the operational databases [3].A key characteristic of data warehousing is that it allows multidimensional view of data. In a data warehouse a multidimensional view is designed and implemented using a star schema or a snow- flake schema. In addition, data warehouses allow a wide range of analytical operations including: *rollup* (increasing the level of aggregation), *drill-down* (decreasing the level of aggregation or increasing detail) along one or more dimension hierarchies, *slicing and dicing* (selection and projection), and *pivoting* (re-orienting the multidimensional view of data) [3].

Later, with the increase of interest in data warehouses, new applications evolved. Data warehouses were then modified to accommodate new types of data including spatial data. Spatial data warehouses were proposed to enable decision making on geographical data. In general, spatial databases store geometric data about objects. These databases are essential in geographical information systems (GIS) as they enhance and utilize geographical applications like demographic analysis. Recently, with the abundance of mobile users and the obvious increase in the number of mobile users worldwide, Spatio-temporal databases and consequently Spatio-temporal data warehouses started to evolve. Spatio-temporal databases deal with geometries changing over time. Clearly, when we try an integration of space and time, we are dealing with geometries changing over time.

In spatial databases, three fundamental abstractions of spatial objects have been identified: A *point* describes an object whose location, but not extent, is relevant, (e.g., a city in a large scale map). A *line* (meaning a curve in space, usually represented as a polyline) describes facilities for moving through space or connections in space (roads, rivers, power lines, etc.). A *region* is the abstraction for an object whose extent is relevant (e.g., a forest or a lake). These terms refer to 2-D space, but the same abstractions are valid in three or higher-dimensional spaces[4]. Considering time as another dimension, spatial-temporal databases are databases about spatial objects that evolve over time. Spatio-temporal objects are also known as moving objects. Moving objects may be points (e.g. cars, buses, planes, mobile users) or regions (e.g. hurricanes, forest fires, etc.). In order to manage, maintain and query historical large amounts of data about moving objects that are stored in Spatio-temporal databases the need for a new technique to deal with this aspect becomes very crucial. Spatio-temporal data warehouse is the technique that is capable to deal and analyze historical spatio-temporal data about moving objects.

Spatio-temporal data warehouses store in it historical information about moving objects trajectories; these trajectories represent the motion of the moving objects. Spatio-temporal data warehouses are also called "trajectory data warehouses" (TDW).

A spatio-temporal data warehouse like any data warehouse has three main phases: extract, transform, and load usually known as "ETL". In[5] the authors contribute with the proposal of the Extract-Transform-Load (ETL) process for reconstructing moving objects, also they addressed the distinct presence measure. In [5, 6] the authors adopt a discrete model for representing moving objects trajectories, they present a technique for counting the distinct number of objects existing in a cell. The proposed technique uses a star schema that involves both spatial and temporal dimensions. The authors also classified the aggregate functions into three groups: algebraic, distributive and holistic. Finally, in [7] the authors presented an approach to compute the algebraic presence measure using an SQL computation approach. The authors also proposed 2 new measures *cross-in* and *cross-out* these measures have a great importance in many location based applications.

Inspired by the importance of spatio-temporal data warehouses, in this paper we aim to present a new dimensional model that computes new measures needed by a wide range of location based applications.

The main contributions of this paper are:

1- Presenting a star schema based dimensional model that captures both the spatial and temporal dimensions.
2- Introducing new measures (facts) like direction majority, number of objects heading from a specific direction (i.e. north or south or east or west direction or combinations of them e.g. Northeast and so on) that are relevant in many moving object queries.
3- Presenting SQL based queries that query about our new proposed measures.

The rest of the paper is organized as follows: section 2 presents a brief summary of related work. Section 3 discusses our proposed work, our proposed schema, our measures and algorithms for computing these measures. Section 4 presents our SQL queries. Section 6 concludes and proposes future work directions.

## 2. RELATED WORK

Nowadays we are witnessing an explosion in the number of mobile phones and the number of mobile users worldwide. In 2011 it was recorded that about 87% of the world's population uses mobile phones and that the number of mobile phones is about 6 billion. This huge number of users raised up the need for new technologies that can efficiently provide services for those users. Moving objects databases (MODs) were thus developed to present efficient management, storage, indexing and querying for location information of those continuously moving objects.

In general, moving objects are objects that change their location and/or shape over time. Moving objects are classified into 2 main categories moving points (e.g. cars, buses, planes, mobile users, etc.), and moving regions (e.g. hurricanes, forests fires, etc.). Being designed to deal with new data, moving objects databases required new techniques that were lately investigated in several research works. Modeling moving objects was one of those issues. Modeling moving objects was introduced in [1, 4, 8, 9] where discrete, continuous, and constraint models were presented. Also, indexing of moving objects is still considered a crucial research direction. This area was explored in many research works as [10-12] where several index structures were proposed. The authors in[12] for example presented a comparison between R Tree, MVR tree and TPR tree as techniques for indexing moving objects. Another key research direction is querying moving objects databases, although several work targeted this direction, it is still an open research area where new query techniques, languages, and algorithms are needed to query moving objects. Several query types were proposed in literature including: range queries, skyline queries, nearest neighbor and reverse nearest neighbor queries [13-15]. Recently, researchers started to explore uncertainty representation and querying in moving objects.

With the vast amount of location information in different applications the need for a technique to manage, maintain and query those data becomes very crucial. Data warehouse is one of the techniques that play a vital role in the decision making process. Data warehouse is defined in [2]as a subject oriented, integrated, time-variant, non-volatile collection of data that helps and

...International Journal of Database Management Systems ( IJDMS ) Vol.5, No.3, June 2013supports the decision making process. Spatial data warehouses are important as they contain large databases with geographical information with a percentage of approximately 80% [16].

Inspired by the importance of spatial data warehouses, several research works studied this area. In [17]the authors proposed a method to perform a selective materialization based on the relative access frequency of the sets of mergeable spatial regions, that is, the sets of mergeable spatial regions should be pre-computed before accessed. In[18] the authors classified the spatial data warehouses' dimensions in to 3 categories: descriptive (Thematic), temporal, and spatial.

Later, spatio-temporal data warehouses emerged. Spatio-temporal data warehouses (also known as trajectory data warehouse) deal with geometries that are changing over time. Spatio-temporal data warehouse was proposed in [6, 19] . New measures were proposed that are tailored to the unique characteristics of moving objects. The "presence" measure is one of the measures that was presented to count the number of objects in a spatial area [6, 19]. The authors adopted a discrete model for representing moving objects trajectories. They presented a star schema with spatial and temporal dimensions also, they classified the aggregate functions into 3 classes: holistic, distributive and algebraic. In [5, 20] the authors studied the presence measure as in [19]. The main contribution of [5] is the proposal of an Extract-Transform-Load process for reconstructing the moving objects. In [21] the authors presented a system that is capable of answering a wide range of traffic related queries also, they proposed a star schema model with spatial and temporal dimensions that captures a number of traffic related measures. Finally in[7] the authors computed the algebraic presence measure using a SQL based computation. The authors also introduced 2 new measures cross-in and cross-out that are useful in many LBSs applications.

## 3. PROPOSED WORK

Following the work on moving object data warehouses, in this paper we continue to investigate the design of spatio-temporal data warehouses with new measures that efficiently analyze the trends and behavior of moving objects. In this paper we focus on moving points. The motion of a moving point is represented by its trajectory which represents the path that the moving objects will take during its motion through time.

### 3.1 Definition 1

A *trajectory segment* is a line segment represented as a pair of triplets$((x_s, y_s, t_s), (x_e, y_e, t_e))$ where $(x_s, y_s, t_s)$and$(x_e, y_e, t_e)$ represent the start and end positions for the segment at times $t_s, t_e$ respectively.

### 3.2 Definition 2

A moving object *trajectory* is defined as a finite sequence of trajectory segments $(s_1, s_2, .., s_i, .., s_n)$. Such that:

1. A moving object trajectory is continuous.
2. For every pair of consecutive segments $s_i, s_{i+1}$ the end point of $s_i$is the start point of $s_{i+1}$.

The trajectory data warehouse is then built on the above frame work. The star schema paradigm is employed as our multidimensional model [2].The schema consists of a fact table and 4 dimensions. The fact table contains of facts (measures) to be evaluated and the foreign keys for the dimensions tables. Those foreign keys together are the composite primary key of the fact table. Our star schema consists of spatial dimensions (*Spatialx* and *spatialy* dimensions), a temporal dimension (Time dimension), and a dimension called *Trajectorysegments* dimension. Figure 1 presents the cube design generated in Microsoft's Analysis Services.

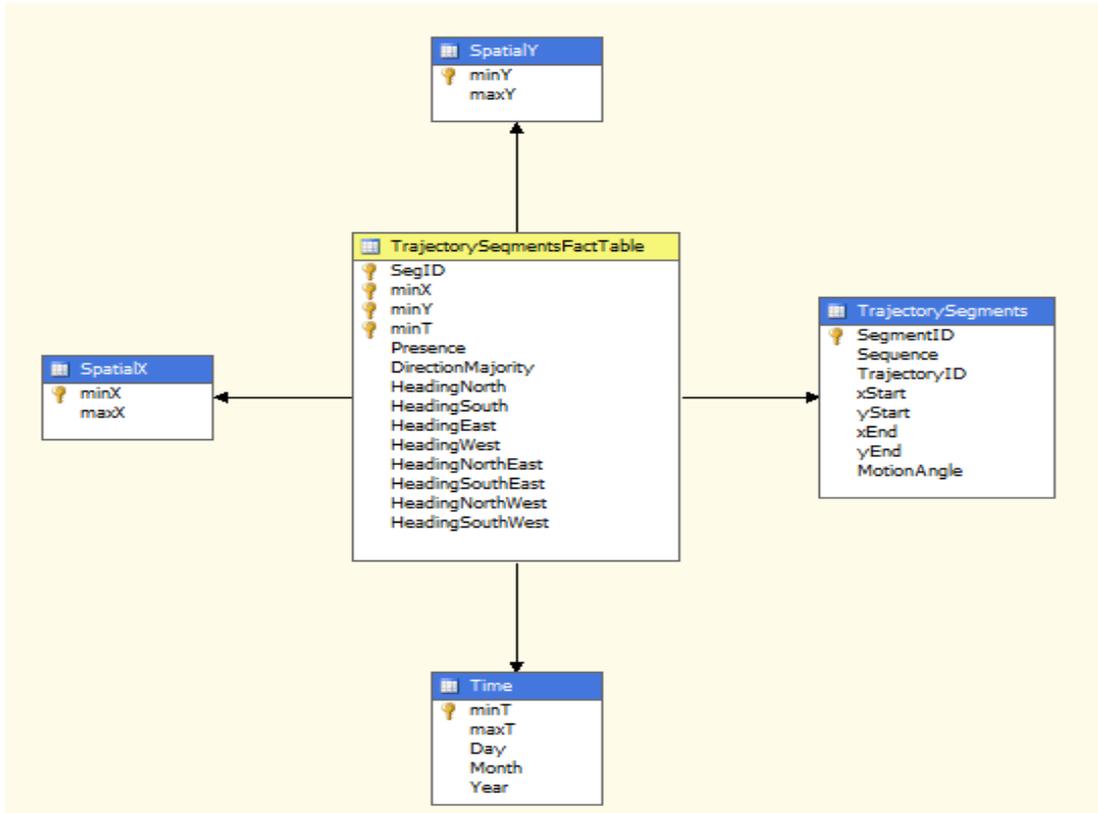

Figure 1.Star Schema

The *SpatialY* dimension has 2 attributes named: minY and maxY. Similarly, the *SpatialX* dimension has 2 attributes named: minX and maxX. The attributes of SpatialY and SpatialX dimensions are used to define the border of a spatial cell. Thus, they that can be used to identify which objects fall within a given cell at a certain time or during a time interval. As proposed in [21], cells can be aggregated to obtain a coarser cell with a larger spatial span. The initial cell size depends on a parameter defined by the user that identifies the grid size based on which the map will be divided. Thus a *spatial hierarchy* is obtained based on grid size. The *Time* dimension contains the traditional time attributes and hierarchies. The hierarchy of the *Time* dimension ranges from day to year.

The final dimension presented in our schema is the *TrajectorySegments* this dimension has the following 8 attributes: *SegmentID* is an identifier for each segment contained in a trajectory moving object, *sequence* identifies the order of each segment in the trajectory, *TrajectoryID* identifies which trajectory does the segment belongs to, (*xStart* and *yStart*) attributes define the start position of each segment, (xEnd and yEnd) attributes define the end position of each segment, and finally the *Angle* attribute is used to define each segment's motion angle (slope).

### 3.3. Definition 3

A *segment motion angle* of segment $s_i$ is the slope of the segment and is computed as:

$$\text{Motion angle } (s_i) = \tan^{-1}\left(\left(\frac{y_{i+1}-y_i}{x_{i+1}-x_i}\right) \times \left(\frac{180}{\pi}\right)\right)$$

The motion angle value ranges between [0°, 360°]. Consequently each segment will eventually belong to one of the following four quadrants: Q1 [0°, 90° [; Q2]90°, 180° [; Q3]180°, 270° [; Q4]270°, 360°].

Hence, each segment will have a direction based on its motion angle. This classification results in a set of 8 possible directions that we refer as the direction domain $(D_d)$ = {N, E, W, S, NE, SE, NW, SW} as shown in Figure 2.

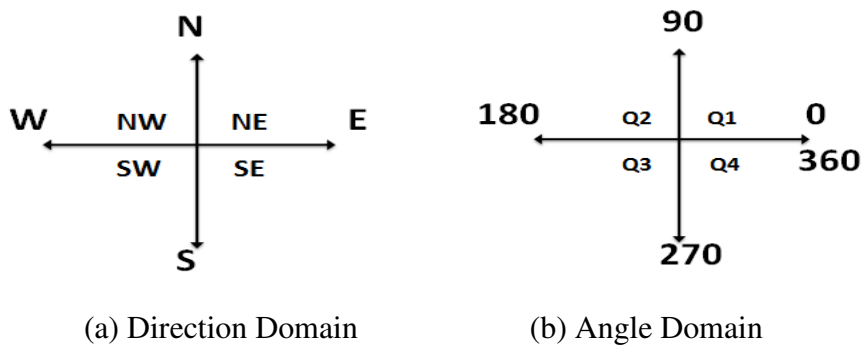

(a) Direction Domain        (b) Angle Domain

Figure 2.Segment's direction domain $(D_d)$ and angle domain

In our proposed trajectory data warehouse schema we introduce measures that are based on motion angle as 1) *DirectionMajority:* Let '$S_c$' be a spatial cell, and 'T' be a time interval, then the *direction majority measure* computes the average motion direction among all the segments existing in $S_c$ during T. 2) *HeadingNorth*: this measure counts the number of objects heading toward the North (N) direction. 3) *HeadingSouth:* this measure counts the number of objects heading toward the South (S) direction. 4) *HeadingNorthEast:* is a measure used to count the number of moving objects heading toward the North East (NE) direction. Similar measures are also proposed for the remaining directions in the direction domain $(D_d)$. Besides, we introduce the presence measure as proposed in [7]. The presence measure is an approximate measure that counts the number of distinct objects existing in the spatial cell $S_c$ during the time interval T. In this paper we continue to compute the presence measure using the same approach presented in [7].

Before feeding the data into the trajectory data warehouse and before constructing the data cube; the data is prepared and pre-processed and this is done in the Extract-Transform-Load (ETL) as in any traditional data warehouse. Our ETL is similar to the one presented in [5, 7] and is implemented using .NET framework (in C# language). Our ETL phase proceeds as follows:

1. We first generate 1000 objects using the Brinkoff generator [22].
2. The trajectory is reconstructed from its sampled data as a set of segments each segment has a start point and an end point.

International Journal of Database Management Systems ( IJDMS ) Vol.5, No.3, June 2013

3. The motion angle value is computed for each segment and inserted in the TrajectorySegments dimension table.
4. Finally, the data is loaded into the data warehouse.

After the completion of the ETL phase the data is ready for analysis and computing the proposed measures.

In the following discussion we present our proposed algorithms for computing the direction-based measures presented above. The first algorithm "Compute_Direction_Count" computes the number of trajectory segments heading toward a certain direction in a spatial cell during a given time interval. Using this count we can identify the direction followed by most objects. The algorithm proceeds as follows: Given a spatial cell, and a set of trajectories$(\tau_1, \tau_2, ...., \tau_n)$. Such that each trajectory is defined as $\tau_i = (s_{i1}, s_{i2}, ..., s_{ik})$, and given a time interval. The algorithm checks the motion angle of each segment in each trajectory and determines the quadrant it falls in. Knowing the quadrant the segments belongs to we identify the motion direction of each segment. Hence, the count of segments moving in this direction is incremented. The details of the algorithm are shown below.

```
Input: Set of Trajectories(τ₁,τ₂,….,τₙ),

Output: Number of trajectory segments in each Direction D ∈ {North,
South, East, West, NorthEast, SouthEast, NorthWest, SouthWest}

ForEachTrajectoryτᵢdo

Begin

ForEachsegment sᵢⱼ ∈ τᵢ , (1≤ j ≤ k) do

Begin

 IF (Motion angle= 0 or Motion angle =360)    Then East ++;

Else IF (Motion angle=90)                     Then North++;

Else IF (Motion angle=180)                    ThenWest++;

Else IF (Motion angle=270)                    ThenSouth++;

Else IF (0<Motion angle<90)                   ThenNorthEast++;

Else IF (90<Motion angle<180)                 ThenNorthWest++;

Else IF (180<Motion angle<270)                ThenSouthWest++;

Else IF (270<Motion angle<360)                ThenSouthEast++;

    End

End
```

Figure3.Compute_Direction_Count algorithm

Using Algorithm 1 we can simply compute the number of objects in each direction. This measure is useful in many traffic related applications where there is a need to determine the route directions that are highly congested. Knowing those congested routes alternative routes could be recommended.

Example 1: Consider the 4 trajectories $\tau_1, \tau_2, \tau_3, \tau_4$ shown below. Assume that the 4 trajectories fall in the same spatial cell, and each trajectories consists of a single segment.

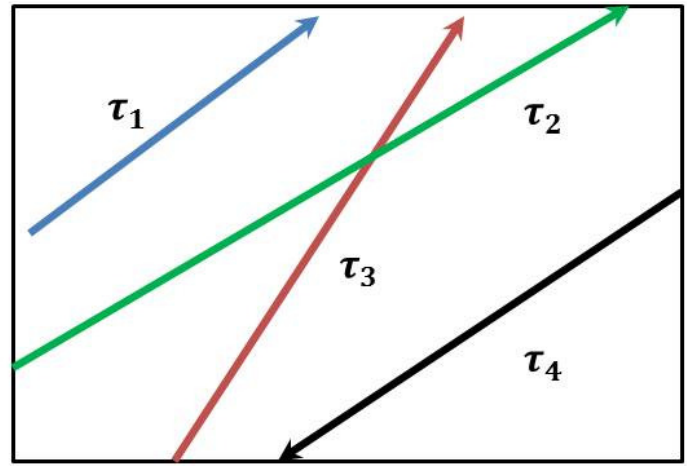

Figure 4. Compute_Direction_Count Example

Using Algorithm1 we can find that trajectories $\tau_1, \tau_2, \tau_3$ are all heading towards the NorthEast (NE) direction. Thus the count for the NE=3. Whereas, trajectory $\tau_4$ is heading towards the SouthWest (SW) direction. Therefore, the count for SW=1. Thus, we can deduce that most of the traffic is heading towards the NE.

In the above example the count was simple as the segment falls completely in the spatial cell. This scenario is not always true, sometimes only portions of the trajectory segment intersect the time interval of interest. In other cases parts of the segment might fall in neighboring cells. In either case the way we compute the direction of the segment should reflect the contribution of the trajectory in that direction. The reason behind this is that we cannot denote a direction as the majority direction used by most of the segments unless the contributions of the segments in that direction is larger than the segments' contribution in other directions.

Algortihm2 presents our proposed approach to compute the segment contribution in each direction. The main idea of the algorithm is as follows: Given a spatial cell, a time interval, and a set of trajectories' segments. For each segment we compute the ratio between the length of the segment portion that falls inside the cell during the time interval in a certain direction and the whole segment length. This ratio is calculated as follows:

$$Segment\ Ratio = \frac{length\ of\ segment\ portion\ inside\ cell}{Total\ segment\ length}$$

Once those ratios are computed for each direction, we then compute the sum along each direction. The direction with the highest value is denoted as the majority direction. The details of the algorithm are shown below.

```
Input: A spatial cell S_c, query interval T, Trajectory τ=(S1,S2,..,Sn).

Output: Direction Majority

ForEach segment s_i in spatial cell S_c, (1≤ i ≤ n) do

  IF (query interval ==cell interval)

    Then

      {
DirectionCount=Compute_Direction_Count (s_i);

Direction Majority=Maximum (DirectionCount);

Return Direction Majority;

      }
 Else

      {

      R= Calculate_Segment_Ratio(s_i);

      /*segment length during time interval/total segment length*/

      Direction Majority= Maximum(R);

      Return Direction Majority;

      }
End
```

Figure 5.Compute_Direction_Majority algorithm

In the above algorithm the Calculate_Segment_Ratio( ) procedure, calculates the contribution of the segment portion in the cell during the time interval in a certain direction. The algorithm is then repeated for each segment in each trajectory. The cumulative sum of ratios in each direction is then computed and the highest valued direction is the majority direction.

Example 2: Consider a trajectory $\tau = (s_1, s_2, s_3)$ with 3 segments. Assume a time window t = [1.8, 3.8] as shown in Figure 6. Assume the start and end positions of each segment are as follows:

$$s_1.start = (1.5, 1.5) \quad s_1.end = (2, 2)$$

$$s_2.start = (2, 2) \quad s_2.end = (3.6, 1.5)$$

$$s_3.start = (3.6, 1.5) \quad s_3.end = (4.5, 2)$$

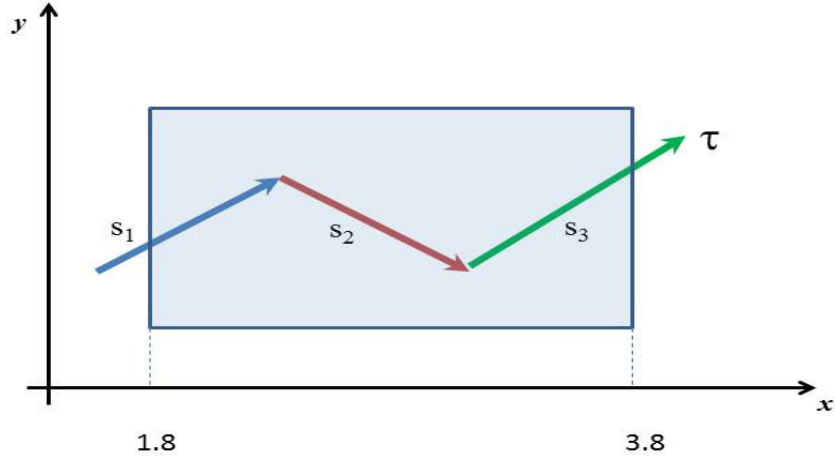

Figure 6. Compute_Direction_Majority Example

To compute the portions of each segment inside the time window we first use linear interpolation to determine the intersection points between the segment and the window borders. Doing so for each segment intersecting the window we get the following interpolated points:

$$s_1.start_{new} = (1.8, 1.8) \quad s_2 \; No \; Changes \quad s_3.end_{new} = (3.8, 1.8)$$

Once we compute those new points we then calculate the length of each segment inside the window, the total segment length, and the segment direction, and we get the following values:

| Segment | Direction | Segment length inside window | Total segment length | Segment Contribution |
|---|---|---|---|---|
| $s_1$ | NorthEast(NE) | 0.3 | 0.71 | 0.4 |
| $s_2$ | SouthEast(SE) | 1.87 | 1.87 | 1 |
| $s_3$ | NorthEast(NE) | 0.2 | 0.9 | 0.22 |

Table1. Direction majority calculations

Thus,

$$\sum Segments \; Contribution \; in \; NE = 0.4 + 0.22 = 0.62$$

International Journal of Database Management Systems ( IJDMS ) Vol.5, No.3, June 2013

Similarly, since $s_2$ falls completely inside the window it contributes to the SE direction with 1.

$$\sum Segments\ Contribution\ in\ SE = 1.0$$

Therefore, we deduce that the majority direction is the SouthEast (SE) direction.

Another important application for the direction-based measures appears in the airfield problems. One of the key factors that affect the airplane performance is the wind direction, gross weight of the plane, atmosphere pressure, runway slope and many others. There is a relation between the wind direction and the runway direction. In general, pilots prefer to take off and land facing into the wind. This has the effect of reducing the aircraft's speed over the ground (for the same given airspeed) and hence reducing the distance required to perform either maneuver. Many airfields have runways facing a variety of directions. The purpose of this is to provide arriving aircraft with the best runway to land on, according to the wind direction. Runway orientation is determined from historical data of the prevailing winds in the area. This is especially important for single-runway airports that do not have the option of a second runway pointed in an alternative direction. Based on the above approach we can benefit from our new measure (Direction Majority) to decide the direction majority of the wind (as a moving object) based on the historical data of the prevailing winds in the area so that the top management can decide whether to have a second runway pointed in an alternative direction or not.

## 4. DIRECTION BASED SQL QUERIES

Once the ETL stage is completed our data warehouse is ready for analysis. Here in this paper we will consider the following measures: presence, DirectionMajority, Count_North, Count_South, Count_East, Count_West, Count_NorthEast, Count_SouthEast, Count_NorthWest, and Count_SouthWest.

The presence measure will be presented as proposed in [7]. These different measures are calculated through the execution of a number of SQL queries (or MDX queries) as follows. (Note that @v1 and @v2 are variables that can be changed to test the query)

### 4.1. Calculating the presence measure

Select Distinct Sum (Presence) AS Total Presence, minX, minY
From TrajectorySegmentsFactTable
Where (minT between @ V1and @v2) and minX=@val1 and minY=@val2
Group by minX, minY

### 4.2. Calculating number of objects moving North

Select Distinct Sum (North) as Total North, minX, minY
From TrajectorySegmentsFactTable
Where (minT between @v1 and @v2) and minX=@val1 and minY=@val2
Group by minX, minY

### 4.3. Calculating Direction_Majority measure

Select DirectionMajority as Direction Majority, minX, minY
From TrajectorySegmentsFactTable
Where (minT between @v1 and @v2) and minX =@val1 and minY= @val2
Group by minX, minY

## 5. CONCLUSIONS AND FUTURE WORK

Spatio-temporal data warehousing (or trajectory data warehouse) continue to be a hot research area. Inspired by the importance of spatial-temporal data warehouses several research works were directed towards designing efficient spatio-temporal data warehouses. In this paper we propose new direction based measures for the trajectory data warehouse that efficiently analyze the trends and behavior of moving objects like direction majority which computes the average motion direction among all segments existing in a spatial cell at a given time interval . We also present SQL queries that are used to compute the proposed measures. For future work we plan to investigate and discover new measures for the trajectory data warehouse and how to apply these new measures in real world application. We also plan to investigate the impact of trajectory uncertainty on spatio-temporal data warehouse models.

### REFERENCES

[1] B. X. O. Wolfson, S. Chamberlain, and L. Jiang, "Moving objects databases: issues and solutions," in *Proc. Int. Conf. on Statistical and Scientific Database Management, 1998, pp. 111–122*.
[2] W. H. Inmon, *Building the Data Warehouse*, 3rd ed. New York, USA: John Wiley & Sons, 2002.
[3] S. C. a. U. Dayal, "An Overview of DataWarehousing and OLAPTechnology" *SIGMOD Record*, vol. 26, pp. 65-74, 1997.
[4] R. H. G. M. Erwig, M. Schneider, and M. Vazirgiannis, "Spatio-temporal data types: an approach to modeling and querying moving objects in databases," *GeoInformatica*, vol. 3, pp. 269-296, 1999.
[5] E. F. G. Marketos, I Ntoutsi, N. Pelekis, A. Raffaeta, and Y. Theodoridis, " Building real world trajectory warehouses," in *Proc. 7th International ACM SIGMOD Workshop on Data Engineering for Wireless and Mobile Access (MobiDE'08)*, Canada, 2008, pp. 8-15.
[6] S. Orlando, Orsini, R., Raffaetà, A., Roncato, A., and Silvestri, C, "Trajectory Data Warehouses: Design and Implementation Issues. ," *Journal of Computing Science and Engineering*, vol. 1, pp. 211–232, 2007.
[7] G. M. H. M. O. Mokhtar "Querying trajectory data warehouses," in DBKDA: *The 1$^{st}$ International Conference on Advances in Databases, Knowledge, and Data Applications*, March 2009, pp. 101–107.
[8] O. W. A. P. Sistla, S. Chamberlain, and S. Dao, "Modeling and querying moving objects," in *Proc. Int. Conf. on Data Engineering*, 1997, pp. 422–432.
[9] R. H. G. Martin Erwig, Markus Schneider, Michalis Vazirgiannis, "Abstract and Discrete Modeling of Spatio-Temporal Data Types.," in *Proceedings of the 6th ACM international symposium on Advances in geographic information systems*, New York, USA, 1998, pp. 131-136.
[10] X. M. Rui Ding, Yun Bai, "Efficient Index Maintenance for Moving Objects with Future Trajectories.," in *Proceedings of the Eighth International Conference on Database Systems for Advanced Applications*, Washington, USA, 2003.
[11] S. S. a. C. S.Jensen, "Indexing of Moving Objects for Location-Based Services," in *Proceedings of the 18th International Conference on Data Engineering,* Washington, USA, 2001, pp. 1-22.
[12] A. Murugappan, "Comparison of R Tree, MVR Tree and TPR Tree."

[13] C. S. J. Rimantas Benetis, Gytis Karciauskas, and Simonas Saltenis, " Nearest neighbor and reverse nearest neighbor queries for moving objects," *The International Journal on Very Large Data Bases*, vol. 15, pp. 229 - 249, 2006.
[14] H. M. O. M. a. J. Su, "Universal trajectory queries for moving object databases," *in IEEE IntC˙onfo˙n Mobile Data Management (MDM'04),* 2004, pp. 133–144.
[15] H. L. B. Zhiyong Huang , Chin Ooi, Anthony K.H. Tung, "Continuous Skyline Queries for Moving Objects " *IEEE Transactions on Knowledge and Data Engineering* vol. 18, pp. 1645-1658, 2006.
[16] M. L. Gonzales, "Seeking spatial intelligence," in *Intelligentent Enterprise*. vol. 3, 2000.
[17] N. S. Jiawei Han, and Krzysztof Koperski, "Selective materialization: An efficient method for spatial data cube construction," in PAKDD '98*: Proceedings of the Second Pacific-Asia Conference on Research and Developmentin Knowledge Discovery and Data Mining*, London, UK, 1998, pp. 144–158.
[18] T. M. Y. Bedard, and J. Han, "Fundamentals of spatial data warehousing for geographic knowledge discovery," in *Geographic Data Mining and Knowledge Discovery*, 2001, pp. 53–73.
[19] R. O. S. Orlando, A. Raffaet, A. Roncato, and C. Silvestr, "Trajectory data warehouses: Design issues and use cases," in *proceedings of the Fifteenth Italian Symposium on Advanced Database Systems SEBD*, 2007, pp. 208–219.
[20] A. R. N. Pelekis, M.L. Damiani, C. Vangenot, G. Marketos, E. Frentzos, I Ntoutsi, and Y. and Theodoridis., " Towards Trajectory Data Warehouses," in *Mobility, Data Mining and Privacy*, 2008, pp. 189-211.
[21] H. M. O. Mokhtar, "HITS: A History-Based Intelligent Transportation System," *International Journal of Data Mining & Knowledge Management Process (IJDKP),* vol. 1, pp. 34-46, 2011.
[22] Brinkhoff, "Generating traffic data," in *IEEE Computer Society Technical Committee on Data Engineering, 2003*, pp. 19-25.

## Authors

Leila Abdelrahman Esheiba

For the time being I am a teaching Assistant in the Information Systems Department, Faculty of Computers and Information, Cairo University. I received my BSc in 2007 from the Information System Department, Faculty of Computers and Information and I was the first of my department. My research interests are Database Systems, Data warehousing and moving objects.

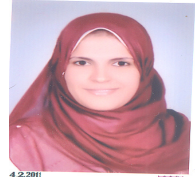